\documentclass[copyright,creativecommons]{eptcs}
\providecommand{\event}{DCM 2010} 
\usepackage{breakurl}             
\usepackage{amsmath, amsthm, amssymb}
\usepackage{graphicx}

\title{Quantum algorithms for testing Boolean functions}
\author{Dominik F. Floess \quad \quad Erika Andersson
\institute{SUPA, School of Engineering and Physical Sciences}
\institute{
Heriot-Watt University, Edinburgh EH14 4AS, United Kingdom 
}
\email{\quad dominikfloess@gmx.de \quad\quad\quad E.Andersson@hw.ac.uk}
\and
Mark Hillery
\institute{Department of Physics, Hunter College of CUNY}
\institute{695 Park Avenue, New York, NY 10061, USA}
\email{mhillery@hunter.cuny.edu}
}
\def\titlerunning{Quantum algorithms for testing Boolean functions}
\def\authorrunning{D. F. Floess, E. Andersson \& M. Hillery}
\begin{document}
\maketitle

\begin{abstract} 
We discuss quantum algorithms, based on the Bernstein-Vazirani algorithm, for finding which variables a Boolean function depends on.
There are $2^n$ possible linear Boolean functions of $n$ variables; given a linear Boolean function, the Bernstein-Vazirani quantum algorithm can deterministically identify which one of these Boolean functions we are given using just one single function query. The same quantum algorithm can also be used to learn which input variables other types of Boolean functions depend on, with a success probability that depends on the form of the Boolean function that is tested, but does not depend on the total number of input variables.
We also outline a procedure to futher amplify the success probability, 
based on another quantum algorithm, the Grover search. 
\end{abstract}

\section{Introduction}

In the oracle identification problem, we are given an oracle from 
a set of possible Boolean oracles, and our task is to determine which
one we have \cite{ambainis1}-\cite{iwama}. The complexity of the problem
is measured by the number of times we must query the oracle in order to
identify it.  Both the Bernstein-Vazirani \cite{bernstein, cleve} and Grover quantum algorithms \cite{grover, groverNM}
solve this type of problem. The Bernstein-Vazirani algorithm identifies linear Boolean functions with a single function query, and Grover's search algorithm finds marked elements in a database with $N$ elements using $\mathcal{O}(\sqrt{N})$ queries.

Consider the following task.  We are given a black box that evaluates a Boolean function
$f(x_{1},x_{2},\ldots x_{n})$ that maps $\{ 0,1 \}^n$ to $\{ 0,1 \}$.  The function depends on the
values of at most $m$ of the variables and is independent of the other $n-m$.  Such a 
Boolean function is called a junta, and, if it depends on only one of the variables, it is called
a dictatorship.  Our task is to find which of the variables the function depends on.  We shall 
show how a variant of the Bernstein-Vazirani algorithm can solve this problem.
Recently, R\"otteler presented a quantum algorithm for identifying quadratic Boolean functions \cite{rotteler}. Atici and Serviedo discuss a quantum algorithm for identifying $k$-juntas, essentially based on the Bernstein-Vazirani oracle \cite{atici}. 
The quantum algorithm we outline is simpler;
moreover, we also present a method to further increase the success probability, based on Grover's quantum search algorithm.

The paper is arranged in the following way. In section \ref{sec:BV}, we review the Bernstein-Vazirani algorithm. In Section \ref{sec:learning}, we show that this quantum algorithm can also be used for the more general task of finding variables other types of Boolean functions depend on. In section \ref{sec:grover}, we show how a method based on the Grover search can be used to further increase the success probability of finding variables the Boolean function depends on. 
We finish with Conclusions.

\section{The Bernstein-Vazirani algorithm}
\label{sec:BV}

The Bernstein-Vazirani algorithm is a one-shot quantum algorithm \cite{bernstein,cleve}.  
It solves the following problem.  One has a black box that evaluates a linear Boolean function,
given by 
\begin{equation}
f(x)=y\cdot x = \sum_{j=1}^{n} y_{j}x_{j} ,
\end{equation}
where the addition is modulo 2 and $y$ is a fixed, but unknown, $n$-bit string.  We want to
find $y$.  The Bernstein-Vazirani algorithm does this with one evaluation of the function.  It
does so by mapping the functions to vectors in an $N$-dimensional Hilbert space, $\mathcal{H}
=\otimes^{n}\mathcal{H}_{2}$, where $N=2^n$ and ${H}_{2}$ is a two-dimensional
Hilbert space.  The computational basis vectors of $\mathcal{H}_{2}$ are $|0\rangle$ and 
$|1\rangle$, and the basis vectors of $\mathcal{H}$ are labeled by $n$-bit strings $|x\rangle
=|x_{1}\rangle \otimes |x_{2}\rangle \ldots \otimes |x_{n}\rangle$.  The function $y\cdot x$ is 
mapped to the vector $v_{y}$, where
\begin{equation}
\langle x|v_{y}\rangle = \frac{1}{\sqrt{N}} (-1)^{y\cdot x} .
\end{equation}
These vectors are orthonormal, i.e.\ $\langle v_{y}|v_{y^{\prime}}\rangle = \delta_{y,y^{\prime}}$,
and they constitute an orthonormal basis of $\mathcal{H}$ known as the parity basis
\cite{bernstein}.  This follows from the identity
\begin{equation}
\sum_{x\in \{ 0,1\}^{n}  } (-1)^{x\cdot y} = \delta_{y,0}  .
\end{equation}
Because the vectors are orthonormal,
they are perfectly distinguishable, and so with one measurement we can perfectly determine
which function the black box is evaluating.

This is actually accomplished by using a circuit consisting of Hadamard gates and an 
$f$-controlled-NOT gate.  The Hadamard gate is the unitary transform
\begin{eqnarray}
|0\rangle &\rightarrow& \frac{1}{\sqrt{2}}(|0\rangle+|1\rangle)\nonumber\\
|1\rangle &\rightarrow& \frac{1}{\sqrt{2}}(|0\rangle-|1\rangle).
\end{eqnarray}
If we apply $n$ Hadamard gates, one to each qubit in the state
$|x\rangle$, we obtain
\begin{equation}
H^{\otimes n}|x\rangle = \frac{1}{\sqrt{N}}\sum_{z\in \{ 0,1\}^{n} } (-1)^{x\cdot z} |z\rangle ,
\end{equation}
where, as before, we have set $N=2^{n}$.  The $f$-controlled-NOT gate, where $f$ is a Boolean
function, acts on $n+1$ qubits in the following way
\begin{equation}
U_{f}|x\rangle |z\rangle = |x\rangle |z+f(x)\rangle ,
\end{equation}
where $|x\rangle$ is an $n$-qubit computational basis sate, $|z\rangle$ is a one qubit state 
($z=0,1$), and the addition is modulo 2.  Now, the input state to the Bernstein-Vazirani circuit 
is the $(n+1)$-qubit state 
\begin{equation}
|\Psi_{in}\rangle  =\frac{1}{\sqrt{2}}|00\ldots 0\rangle (|0\rangle - |1\rangle ) .
\end{equation}
We first apply $n$ Hadamard gates, one to each of the first $n$ qubits, and then the 
$f$-controlled-NOT gate, giving us
\begin{equation}
|\Psi_{in}\rangle \rightarrow \frac{1}{\sqrt{2N}} \sum_{x\in \{ 0,1\}^{n}  } (-1)^{f(x)}|x\rangle
(|0\rangle - |1\rangle )
\end{equation}
Next, we again apply $n$ Hadamard gates to the first $n$ qubits yielding 
\begin{equation}
|\Psi_{out}\rangle = \frac{1}{N\sqrt{2}}  \sum_{x\in \{ 0,1\}^{n}  } \sum_{z\in \{ 0,1\}^{n}  }
(-1)^{f(x)+x\cdot z}|z\rangle (|0\rangle - |1\rangle ) .
\end{equation}
Discarding the last qubit (it is not entangled with the others, so this has no effect) and expressing
this result in terms of the vectors $|v_{y}\rangle$, we find the $n$-qubit output state
\begin{equation}
|\psi_{out}\rangle = \sum_{z\in \{ 0,1\}^{n}  } \langle v_{z}|v_{f}\rangle |z\rangle ,
\end{equation}
where we have defined the vector $v_{f}$ to have the components $\langle x|v_{f}\rangle = 
(1/\sqrt{N}) (-1)^{f(x)}$.  Now, if we know that $f(x)$ is of the form $f(x)= y\cdot x$, then we just 
get the vector $|y\rangle$ as our output, and when we measure $|\psi_{out}\rangle$ in
the computational basis, we find the $n$-bit string $y$.  Therefore, we find out what the function 
is with only one application of the $f$-controlled-NOT gate.  Classically, we would need to 
evaluate the function $n$ times to find $y$.

\section{Learning which variables a general Boolean function depends on}
\label{sec:learning}

If $f(x)$ is a general Boolean function, then when we measure $|\psi_{out}\rangle$ in the 
computational basis, we will obtain the label of one of the basis vectors $v_{y}$, with
which $v_{f}$ has a nonzero overlap.  The key to using this to solve the problem stated
in the Introduction is the following fact: if $f(x_{1},x_{2},\ldots x_{n})$ is independent of 
the variable $x_{j}$, and $y\in \{ 0,1\}^{n}$ has the property that $y_{j}=1$, then
$\langle v_{y}|v_{f}\rangle = 0$.  In order to prove this we start by noting
\begin{eqnarray}
\langle v_{y}|v_{f}\rangle & = &  \frac{1}{\sqrt{N}} \sum_{x\in \{ 0,1\}^{n}  } (-1)^{f(x)+x\cdot y}
\nonumber \\
 & = &  \frac{1}{\sqrt{N}}\sum_{x_{1}=0}^{1} \ldots \sum_{x_{n}=0}^{1} (-1)^{f(x)+x\cdot y} .
\end{eqnarray}
Now look at the $x_{j}$ sum,
\begin{equation}
\sum_{x_{1}=0}^{1} (-1)^{f(x)+x\cdot y}  = (-1)^{f(x)} \prod_{k=1, k\neq j}^{n} (-1)^{x_{k}y_{k}}
\sum_{x_{j}=0}^{1} (-1)^{x_{j}} = 0 .
\end{equation}
This proves our result.  What it implies is that if we use the Bernstein-Vazirani circuit with a
Boolean function that is a junta and find an output vector $|y\rangle$ that has ones in a number
of places, then the function does depend on the variables corresponding to those places. If the function does not depend on a particular input variable, then the $n$-qubit state $|\psi_{out}\rangle$ will always have a 0 in that position.

It is important to note that the success probability to find the variables the function depends on is {\it  independent} of the total number $n$ of input variables. 
In general, the success probability for the quantum algorithm depends only on the form of the Boolean function that is being tested, that is, it depends on the number of significant variables, and the functional form of the Boolean function involving these significant variables.

\subsection{Boolean functions depending on only two input variables}

In order to illustrate this, let us consider a simple example.  Suppose that we know that our function is
given by $f(x_{1},x_{2},\ldots x_{n})=x_{j}x_{k}$, but we do not know $j$ and $k$, i.e. we know that
the Boolean function is the product of two of the variables, but we do not know which two.
Our task is to find out which two.  The vector $|v_{f}\rangle$ corresponding to this function has
a nonzero inner product with only four of the basis vectors $|v_{y}\rangle$.  We must have $y_{l}=0$
for $l\neq j,k$, which leaves four possibilities, which we shall denote by $|y_{00}\rangle$,
corresponding to $y_{j}=y_{k}=0$, $|y_{01}\rangle$, corresponding to  $y_{j}=0$ and $y_{k}=1$,
etc.  We find that the output of the Bernstein-Vazirani circuit in this case is
\begin{equation}
|\psi_{out}\rangle = \frac{1}{2}(|y_{00}\rangle + |y_{01}\rangle + |y_{10}\rangle - |y_{11}\rangle ).
\end{equation}
If we measure in the computational basis, then we will obtain one of these basis 
vectors.  If we obtain $|y_{00}\rangle$, we learn nothing, and the procedure has failed.  This
happens with a probability of $1/4$.  If we obtain either $|y_{01}\rangle$ or $|y_{10}\rangle$,
then we learn one of the variables, and if we obtain $|y_{11}\rangle$, we obtain both.  All of
these outcomes have a probability of $1/4$, so that we learn at least one of the variables on
which the function depends with a probability of $3/4$. This probability is independent of how many input variables $n$ there are in total.
Classically, a 
possible procedure would be to initially set all of
the variables equal to $1$, which would set the value of the function equal to $1$.  One then
changes the value of the variables, one at a time, to see which ones cause the value of the
function to change.  In order to learn on which variables the function depends, one would
have to evaluate the function $\mathcal{O}(n)$ times.  If $n$ is large, the quantum procedure,
though probabilistic, is more efficient. 

Let us now consider a somewhat more general example.  We will still assume that our function only depends
on two out of the $n$ variables, $x_{j}$ and $x_{k}$, but we will not assume the specific form 
of the function.  We can express $f(x_{1},x_{2},\ldots x_{n})$ as
\begin{equation}
f(x_{1},x_{2},\ldots x_{n}) = g(x_{j},x_{k}) ,
\end{equation}
where $g(x_{j},x_{k})$ is some Boolean function of two variables.  Now, assuming that $y_{l}=0$
for $l\neq j,k$, we have that
\begin{equation}
\langle v_{f}|v_{y}\rangle = \frac{1}{4} \sum_{x_{j},x_{k}=0}^{1} 
(-1)^{g(x_{j},x_{k})+y_{j}x_{j}+y_{k}x_{k}} .
\end{equation}
The right-hand side of the equation can only be $0$, $1$, or $\pm 1/2$, and it will only be $0$
or $1$ if $f(x_{1},x_{2},\ldots x_{n})$ is one of the basis functions.  Therefore, $|\psi_{out}\rangle$
is either one of the vectors $|y_{l_{1}l_{2}}\rangle$, or of the form
\begin{equation}
|\psi_{out}\rangle = \frac{1}{2}(\pm |y_{00}\rangle \pm |y_{01}\rangle \pm |y_{10}\rangle 
\pm |y_{11}\rangle ).
\end{equation}
If $f(x_{1},x_{2},\ldots x_{n})$ is one of the basis functions, we succeed after one trial, however, 
we do not know this, and several trials in which we get the same answer will be necessary to
confirm that we have one of the basis functions.  In all other cases, we will fail, that is get no
information about which variables the function depends on, with a probability of $1/4$, so that
after several trials we will, with high probability, know $x_{j}$ and $x_{k}$.

\subsection{Boolean functions depending on more than two input variables: an example}

Now let us see what happens if the function depends on more than two variables.  We know that the quantum algorithm will always find the variables a function depends on, but that the success probability for this will vary with the form of the Boolean function. Let us
consider the case
\begin{equation}
f(x_{1},x_{2},\ldots x_{n}) = \prod_{j=1}^{m}x_{j}  .
\end{equation}
The probability to identify which variables this function depends on would be the same also for other Boolean functions which are a product of any $m$ out of the $n$ variables.
For vectors $|v_{y}\rangle$ such that $y_{j}=0$ for $j>m$, we have
\begin{equation}
\label{mproduct}
\langle v_{f}|v_{y}\rangle = \frac{1}{2^{m}} \sum_{x_{1}=0}^{1} \ldots \sum_{x_{m}=0}^{1}
(-1)^{h(x_{1}, \ldots x_{m}; y)} ,
\end{equation}
where 
\begin{equation}
h(x_{1},\ldots x_{m}; y) = \prod_{j=1}^{m}x_{j} + \sum_{j=1}^{m}x_{j}y_{j}  .
\end{equation}
Now, if the product $x_{1}x_{2}\ldots x_{m}$ were absent from the exponent in Eq.\ (\ref{mproduct}),
and if at least one of the $y_{j}\neq 0$, then the sum would be zero.  The product changes the
sign of only one of the terms, so that we have
\begin{equation}
\langle v_{f}|v_{y}\rangle = \pm \frac{1}{2^{m-1}} .
\end{equation}
If $y_{j}=0$ for $j=1, \dots n$ (we shall denote the vector corresponding to this $y$ by 
$|v_{0}\rangle$), then without the product in the exponent all of the terms in the sum in Eq.\ 
(\ref{mproduct}) would be $1$.  The presence of the product again changes only one term,
so that
\begin{equation}
\langle v_{f}|v_{0}\rangle = 1 -  \frac{1}{2^{m-1}} .
\end{equation}
Note that since the failure probability is just $|\langle v_{f}|v_{0}\rangle |^{2}$, this implies that
the failure probability grows with $m$. 
This is the ``worst case scenario"; this type of Boolean function belongs to the class of functions for which the Bernstein-Vazirani algorithm has least probability to succeed in finding the variables it depends on, since a phase factor is added only to a single term. Nevertheless, the success probability is still independent of the total number $n$ of input variables.

\section{Amplification of the success probability}
\label{sec:grover}

The desirable outcomes of the measurement of the output state $|\psi_{out}\rangle$ are those with as many 1's as possible, since a ``1" in position $i$ indicates that the Boolean function depends on input variable $x_i$.  To further increase the success probability,  
it is possible to amplify components of $|\psi_{out}\rangle$ with a chosen number and above of 1's.
This procedure is based on Grover's quantum search algorithm. 
Grover's algorithm uses $\mathcal{O}(N/M)$ queries for searching a database with $N$ elements, where $M$ of these are solutions to the search problem \cite{grover, groverNM}. Classically, $\mathcal{O}(N/M)$ database queries are needed. 

Let us define the normalised states $|\alpha\rangle$ and $|\beta\rangle$ as
\begin{eqnarray}
|\alpha\rangle &=& A {\sum_x}^{''} v_x |x\rangle; \quad A = \frac{1}{\sqrt{\sum^{''}v_x^2}} \\
|\beta\rangle  &=& B {\sum_x}^{' } v_x |x\rangle;   \quad B = \frac{1}{\sqrt{\sum^{'}v_x^2}},
\label{eqn:alpha-betha}
\end{eqnarray}
where the prime $'$ indicates a sum over all $x\in\{0,1\}^n$ which contain $k$ or more 1's and $''$ indicates a sum over the remaining $x$.
The state $|\psi_{out}\rangle$ in terms of $|\alpha\rangle$ and $|\beta\rangle$ is
\begin{equation}
|\psi_{out}\rangle = \frac 1 A |\alpha\rangle + \frac 1 B |\beta\rangle = \cos \frac \theta 2 |\alpha\rangle + \sin \frac \theta 2 |\beta\rangle,
\end{equation}
where $\cos\theta = 1/A=\sqrt{\sum^{''}v_x^2}$ and $\sin\theta=1/B=\sqrt{\sum^{'}v_x^2}$. 
Repeated application of the operator
\begin{equation}
G=H^{\otimes n} U_fH^{\otimes n} (2|0\rangle\langle 0|-{\bf 1})H^{\otimes n}U_fH^{\otimes n}O,
\end{equation}
where the operator $O$ produces phase factors $-1$ for components with $k$ or more 1's, gives 
\begin{equation}
G^l |\psi_{out}\rangle = \cos \left( \frac{2l+1}{2}\theta \right) |\alpha\rangle + \sin  \left( \frac{2l+1}{2}\theta \right) |\beta\rangle .
\end{equation}
after $l$ applications. The optimal number of Grover iterations is given by the integer closest to
\begin{equation}
R(\gamma) = \frac{\arccos [\sin(\theta/2)]}{\theta}=\frac{\arccos\sqrt{\gamma}}{2\arcsin\sqrt{\gamma}},
\end{equation}
where $\gamma=\sum^{'}v_x^2$. The leading term in the power series expansion of $R(\gamma)$ about $\gamma=0$ is $\pi/(4\sqrt{\gamma})$. All higher order terms have a negative sign. Hence we have
\begin{equation}
R < \frac \pi {4\sqrt \gamma},
\end{equation}
and if $\gamma \ll 1$, then
\begin{equation}
R \lesssim \frac \pi {4\sqrt \gamma}. 
\end{equation}
For this number of iterations, the final state contains the largest possible fraction of the component $|\beta\rangle$.
If the form of the Boolean function is known (e.g. that the Boolean function is of the form $x_i x_j$, but not what $i,j$ are), then it is possible to calculate $\gamma$ and the optimal number of Grover iterations for the chosen value of $k$. The smaller $k$ is chosen, the larger $\gamma$ is, and the fewer Grover iterations are needed. If the form of the function is not known, then, just as for the usual Grover search algorithm, it is possible to estimate the optimal number of Grover steps \cite{dominikthesis}. This will require more queries of the function to be tested. It does, however, not necessarily mean that a significantly greater number of function queries is needed; this is the case for the example below.

\subsection{Amplification for a single term of order $k$}

As an example, let us consider the case where $f(x_{1},x_{2},\ldots x_{n}) = \prod_{j=1}^{m}x_{j} $, and suppose that we want to identify all variables this function depends on. As also pointed out before, the success probability would remain the same for any Boolean function which is a product of $m$ input variables. From equation (20), we obtain $\gamma = 2^{-2m+2}$, and consequently the optimal number of Grover iterations needed in order to obtain a high probability of identifying all input variables the function depends on is given by the integer closest to $R=\pi~2^{m-3}$, which is $\mathcal{O}(2^m)$. Each iteration uses two queries of the Boolean function, so that the total number of function queries is roughly $2R=\pi~2^{m-2}$, which is also $\mathcal{O}(2^m)$. We point out that this number is independent of $n$, which is the total number of input variables.

If the Boolean function is a product of $m$ of the input variables, but we do not know this, then we first need to estimate the optimal number of Grover iterations. It can be shown \cite{dominikthesis} that for a product of $m$ input variables, the circuit for estimating the optimal number of Grover steps requires $\mathcal{O}(2^m)$ function queries. In other words, if we are looking to amplify terms with $m$ or more 1's, that is, to find all variable the function depends on, then having to estimate the required number of Grover iterations does not change the order of how many function queries are needed in total.

We can compare the success probability of the amplification strategy to the case where we run the unmodified Bernstein-Vazirani algorithm roughly $2R=\pi~2^{m-2}$ times (the number of runs is given by the integer closest to this number). In each round, the failure probability is $(1-2^{-m+1})^2$, so that the probability to fail in all rounds, learning none of the variables the function depends on,  is approximately $p_f=(1-2^{-m+1})^{\pi~ 2^{m-1}}$. The probability to obtain at least one variable is therefore approximately $1-p_f$, which approaches $1-e^{-\pi}\approx 0.96$ when $m$ becomes large. On the other hand, the probability to never learn one particular variable $x_i$ that the function  does depend on, in any of the $2R=\pi~2^{m-2}$ tries, is equal to
\begin{equation}
p(\text{not learn } x_i) = \left(\sum_{v_y:y_i=0}|\langle v_f|v_y\rangle|^2\right)^{\pi 2^{m-2}} = (1-2^{-m+1})^{\pi 2^{m-2}}.
\end{equation}
This probability approaches $e^{-\pi/2}\approx 0.21$ when $m$ becomes large. For $2R$ function queries, there is therefore an appreciable probability for not learning at least one variable the function depends on when using the Bernstein-Vazirani algorithm without amplification. 
The amplified procedure is very likely to obtain \emph{all} variables which the function depends on with a similar number of function queries. Amplitude amplification for terms with $m$ 1's has therefore improved the situation. 

\section{Conclusions}
We have shown that the Bernstein-Vazirani algorithm may be used for testing which input variables  an unknown Boolean function depends on. In a sense, this task is more general than distinguishing between linear Boolean functions, which is the task for which the Bernstein-Vazirani algorithm was originally devised. The success probability of finding variables a Boolean function depends on may be further enhanced by an amplification procedure based on Grover's search algorithm.

The success probability for the presented quantum algorithm depends on the particular form of the Boolean function, but has the general property that it is independent of the total number of input variables. It shares this property with the algorithm presented in \cite{atici}. Nevertheless, a full comparison of the success probabilities of the different quantum and classical algorithms remains to be made. Other variations of the Bernstein-Vazirani algorithm may also be tailored for investigating Boolean functions of particular forms, and this will be the subject of further investigations.

\bibliographystyle{eptcs} 

\end{document}